\documentclass{aa}
\usepackage{amssymb}
\usepackage{amsfonts}
\usepackage{amsmath}
\usepackage{textcomp}
\usepackage{latexsym}
\usepackage{array}
\usepackage{rotating}

\usepackage{graphicx}
\usepackage{natbib}
\usepackage{txfonts}

\def\apjl{ApJ~Lett.}          
\def\apjs{ApJS}               
\def\aap{A\&A}                

\newcommand{\fv}[1]{\frac{\delta #1}{#1}}
\newcommand{\subrats}{\begin{turn}{36}\!$_{\star}$\end{turn}}

\newcommand{\swapind}{\ensuremath{(1 \Leftrightarrow 2)}}

\newcommand\beq{\begin{equation}}
\newcommand\eeq{\end{equation}}

\begin{document}
   \title{Orbital eccentricities of binary systems with a former AGB star}

   \author{A.A. Bona\v{c}i\'c Marinovi\'c, E. Glebbeek
    \and O.R. Pols
          }
   \authorrunning{Bona\v{c}i\'c Marinovi\'c et al.}
   \offprints{A. Bona\v{c}i\'c Marinovi\'c}

   \institute{Sterrekundig Instituut Utrecht (SIU), Universiteit Utrecht,
     P.O. Box 80000, NL-3508 TA Utrecht, The Netherlands.\\
              \email{bonacic@astro.uu.nl}
             }

   \date{Received July 17, 2007; accepted October 24, 2007}


  \abstract
      {Many binary stellar systems in which the primary star is
	beyond the asymptotic giant branch (AGB) evolutionary
	phase show significant orbital
	eccentricities whereas current binary interaction models
	predict their orbits to be circularised.}
      {In the search for a mechanism to counteract the circularising
	effect of tidal interaction we analyse how the orbital
	parameters in a	system are modified under mass loss and mass
	exchange among its binary components.}
      {We propose a model for enhanced mass-loss from the AGB star
	due to tidal interaction with its companion,
	which allows a smooth transition between the wind and Roche-lobe
	overflow mass-loss regimes. We explicitly follow its
	effect along the orbit on the change of eccentricity and
	orbital semi-major axis, as well as the effect of accretion
	by the companion. We calculate timescales for the variation
	of these orbital parameters and compare them to the tidal
	circularisation timescale.}
      {We find that in many cases, due to
	the enhanced mass loss of the AGB component at orbital
	phases closer to the periastron, the net eccentricity growth
	rate in one orbit is comparable to the rate of tidal circularisation.
	We show that with this eccentricity enhancing mechanism it is
	possible to reproduce the orbital period and eccentricity
	of Sirius system, which under the standard assumptions of binary
	interaction is
	expected to be circularised.
	We also show that this mechanism may
	provide an explanation for the eccentricities of most barium star
	systems, which are expected to be circularised due to tidal
	dissipation.
      }
      {By proposing a tidally enhanced model of mass loss from AGB
	stars we find a mechanism which efficiently works against
	the tidal circularisation of the orbit. This mechanism can
	explain the significant eccentricities observed in binary
	systems containing a white dwarf and a less
	evolved companion, which are predicted to be circularised
	due to their proximity,
	such as Sirius and systems with barium stars.}


   \maketitle
%
\section{Introduction}
%

Detached binary systems containing a white dwarf and a
relatively unevolved companion, i.e., a main-sequence star
or a (sub)giant, are a useful
tool to understand the binary evolution of systems with an
asymptotic giant branch (AGB) star, given that
their orbital and chemical properties do not change significantly
from the moment that the primary finished its AGB evolution
and became the current white dwarf.
An example of such a system is Sirius, a 2.1 M$_{\odot}$
main sequence star with a 1.05 M$_{\odot}$ white dwarf companion
in a 50-year orbit
\citep[e.g., ][]{1960JO.....43..145V,1978ApJ...225..191G}.
Under the standard picture of binary evolution with tidal
interaction this system should have circularised when the primary
became an AGB star, as we show in $\S$\ref{sec:sirius} of this paper.
However,
this system has an eccentricity $e=0.59$ and is not an exception.
A similar problem is faced when considering the barium-star systems,
which are
red giants with over-abundances of $s$-process elements
(prominently barium) with white dwarf companions.
The best current explanation for the enhancement of
$s$-process elements in these stars is that they accreted
mass from
their companion when it was an AGB star. Due to the large size
of AGB stars these systems are expected
to be circularised by tidal interaction for periods smaller
than about 3500-4000 days \citep{2003ASPC..303..290P}.
However, barium-star systems with period as short as 600 days
are observed to be significantly eccentric
\citep{1998A&A...332..877J}. 
The problems stated above are indications that a mechanism
must exist that counteracts the circularising effect of the
tides by the time the primary star is on the AGB.
\citet{1995A&A...293L..25V} propose that
enhanced mass loss at periastron could enhance the
eccentricity and \citet{2000A&A...357..557S}
shows that this works when the AGB star fills its Roche lobe
during periastron passages.
It has also been proposed that the tidal interaction between a
binary system and a circumbinary disk could account for an
eccentricity enhancing mechanism
\citep{1996A&A...314L..17W,1998Natur.391..868W}. However,
\citet{2000A&A...357..557S} argues that
the masses of observed circumbinary disks are too small to
counteract the circularisation.

Mass loss from stars in binary systems is commonly treated as
single-star wind mass-loss as long as the stars are inside their Roche
lobes, while when one star fills its Roche lobe
the mass-loss rate increases abruptly
to the high values that correspond to Roche-lobe overflow.
This approximation is fairly accurate for stars with a steep density
gradient in their atmospheres, but it is not appropriate
for the case of AGB stars, given 
their large atmospheric pressure scale height and their
weakly bound envelope.
The fact that AGB stars undergo dynamical pulsations further reduces
density gradient in the layers above the photosphere
\citep[e.g.,][]{1988ApJ...329..299B}.
We propose a prescription for
enhanced mass loss which
smoothly grows from the single star wind mass loss rate to the Roche
lobe overflow mass loss rate as the radius of the star approaches
its Roche lobe radius. This gives a variable
mass loss rate along eccentric orbits which is higher
at orbital phases closer to the periastron, even if the star
does not fill its Roche lobe. This effect works in the same way
as discussed above,
but no filling
of the Roche lobe is needed, so that (during the AGB phase)
this effect is permanently competing against tidal circularisation.

\citet{1988A&A...205..155B} carried out calculations of the
variation of orbital parameters considering
instantaneous mass transfer and only linear momentum conservation.
Angular momentum conservation was not taken into account in this
exploratory study.
An extension including angular momentum conservation 
was made by \citet{2000A&A...363..660L}, but
they make use of the orbital average of the distance between the
components and orbital angular velocity to carry
out their calculations, which in the case of a
variable mass-loss along the orbit does not give the correct
results.

In $\S$\ref{sec1} we revise the variation of orbital parameters
due to stellar mass loss and mass transfer by taking into account
the conservation laws of linear and angular momentum with respect
to the centre of mass of the actual binary system. We do not make
use of the orbital averages a priori, but give the variations as a
function of the orbital phase, allowing different rates
of mass loss and mass transfer along the orbit.
In $\S$\ref{sec2} we calculate the rate of change of the eccentricity
according to different assumptions of mass loss and accretion.
In $\S$\ref{sec3} we evaluate
the competition between the tidal circularisation and the eccentricity
pumping due to our proposed mass loss, and show that the latter can
prove effective in counteracting the former for systems such as
Sirius and barium stars in which their AGB
component did not fill its Roche lobe. A summary is presented in
$\S$\ref{sec4}.

\section{Variational analysis of orbital parameters due to wind mass
transfer or mass loss}
\label{sec1}
The dynamics of a binary orbit are well
described by the two masses, $M_1$ and $M_2$, its energy,
$E_{\rm orb}$, and its angular momentum, $J_{\rm orb}$
\citep[e.g.,~][]{1980clme.book.....G}. The
fractional variation of the semi-major axis (also called the separation),
$a$, and of the eccentricity, $e$, are expressed as
\begin{equation}
\label{fva}
\fv{a}=\fv{M}+\fv{m}-\fv{E_{\rm orb}}
\end{equation}
and
\begin{equation}
\label{fve}
\fv{(1-e^2)}=2\fv{J_{\rm orb}}+\fv{E_{\rm orb}}-2\fv{M}-3\fv{m},
\end{equation}
where we use the notation for the reduced mass
problem, such that the total mass $M\equiv M_1+M_2$ and the reduced
mass $m\equiv\frac{M_1M_2}{M_1+M_2}$. For a detailed derivation of
these expressions and those that follow in this section
see Appendix \ref{appa}.

\subsection{Angular momentum and energy}
In a system where $r$ is the instantaneous distance between the stars
and $\omega$ is the instantaneous angular velocity of the components with
respect to a co-moving inertial frame with its origin in the centre of mass
(from here on the CM frame),
the fractional variations of orbital angular momentum and energy are
given by
\begin{equation}
\label{fvJ}
\begin{split}
\fv{J_{\rm orb}}&=\fv{m}+2\fv{r}+\fv{\omega}\\
&=\fv{m}+\fv{r}+\fv{v_{\theta}}
\end{split}
\end{equation}
and
\begin{equation}
\label{fvE}
\fv{E_{\rm orb}}=\fv{m}+\frac{2a}{r}\fv{M}-\frac{2a}{r}\fv{r}-
\frac{a}{GM}
\left[2v_r\delta v_r+2v_{\theta}\delta v_{\theta} \right]
\end{equation}
respectively. We denote the velocity of one star
relative to its companion by $\vec{v}=\vec{v_1}-\vec{v_2}$,
where $\vec{v_i}$ is the velocity of star $i$ with respect to the
CM. $\vec{v}$ also corresponds to the orbital velocity in the reduced
mass problem. We also use the subscripts $r$
and $\theta$ to indicate the radial and
transverse velocity components respectively.

We consider a system in which both stars may lose mass in isotropic
winds, part of which may be accreted by the companion such that during
a time interval $\delta t$ an amount of mass $\delta M_{1,{\rm W}}$
is lost by star $1$ and an amount $\delta M_{2,{\rm ACC}}$ is accreted
by star $2$. In this way an amount of mass
$\delta M_{1,{\rm LOST}}=\delta M_{1,{\rm W}}-\delta M_{2,{\rm ACC}}$
is effectively lost by star $2$ from the system.
To determine how the orbital velocity components vary we assume that
star $2$ receives an impulse when it accretes an amount of mass
$\delta M_{2,{\rm ACC}}$ coming from the wind of star $1$, which at
the moment it is accreted has a velocity $\vec{w_{12}}$ in the CM frame.
The corresponding expressions for mass lost from star $2$ are obtained by 
interchanging the indices $1$ and $2$.
This yields the variations
\begin{equation}
\label{dvr}
\delta v_{r}=\big[w_{r,12}-v_{r,2}\big]
\frac{\delta M_{2,{\rm ACC}}}{M_2} + \swapind,
\end{equation}
and
\begin{equation}
\label{dvt}
\delta v_{\theta}=
\big[w_{\theta,21}-r_2\omega\big]\frac{\delta M_{2,{\rm ACC}}}{M_2} + \swapind
\end{equation}
where $r_i$ corresponds to the distances of star $i$ to the CM. The notation
$ + \swapind$ used here and below means: add terms to the left again with
indices $1$ and $2$ interchanged. For quantities defined with an index in the
left-hand side of the equation it is implicit that the corresponding expression
for the other star is obtained by exchanging the indices in the right-hand side.
By means of Eqs. (\ref{fvJ}), (\ref{fvE}), (\ref{dvr}) and (\ref{dvt})
we can now write the orbital angular momentum and energy
variations as
\begin{equation}
\label{fvJorbvect}
\begin{split}
\fv{J_{\rm orb}}=&-\frac{\delta M_{\rm 1,LOST}}{M_1}
\left(\frac{M_2}{M}\right)\\
&+\frac{\delta M_{\rm 2,ACC}}{M_2}
\left(1-\frac{M_2}{M_1}+\frac{w_{\theta{\rm rel},2}}
{r\omega}\right) + \swapind
\end{split}
\end{equation}
and
\begin{equation}
\label{fvEvect}
\begin{split}
\fv{E_{\rm orb}}=
&-\frac{\delta M_{1,{\rm LOST}}}{M_1}\left[1+\left(
\frac{2a}{r}-1\right)\frac{M_1}{M}\right]\\
&+\frac{\delta M_{2,{\rm ACC}}}{M_2}\left[
1-\frac{M_2}{M_1}-\frac{2a}{GM}~\vec{v}\cdot
\vec{w_{{\rm rel},2}}
\right]+\swapind.\\
\end{split}
\end{equation}
In these equations
$\vec{w_{{\rm rel},2}}\equiv(\vec{w_{12}}-\vec{v_2})$ is the
relative velocity of the wind with respect to star $2$ when accreted.
We have also used the fact that there must be no displacements
$\delta r_i$, which implies that $\delta r/r=0$
(see Appendix \ref{appa} for details).

\subsection{Behaviour of the wind}
\label{wbehave}
To apply these expressions in calculations of orbital evolution,
the detailed behaviour of the wind as it travels through the system
must be followed from its point of origin until it is accreted or
escapes the binary potential well. In general this requires either
ballistic calculations \citep[e.g.,][]{1993ApJ...410..719B}
or full-scale hydrodynamical
calculations \citep[e.g.,][]{1996MNRAS.280.1264T,2004A&A...419..335N}.
For the purpose of this paper we assume the
simple model described below.

The velocity of the wind from star $j$ when it is accreted by star
$i$ has a radial component, $w_{r,ji}$, that we will assume
to be the velocity of the wind emanating isotropically from star
$j$ as if it were a single star, $w_j$.
For the transverse component, $w_{\theta,ji}$,
we assume that the specific angular momentum of the wind with respect to
the CM is conserved and we express it as
$h_{{\rm W},j}\equiv\chi_j{r_j}^2\omega$. Note that
the wind carries both the orbital and spin angular momentum of the
mass-losing star. With the above definition
$\chi_j=1$ represents a point mass star, given that such an idealised
object can only carry orbital angular momentum and no
intrinsic spin. Once the wind is accreted
by the companion star $i$ a fraction $\mu_i$ of $h_{{\rm W},j}$ is
transfered to the spin angular momentum of the accretor and the rest
is transfered to the orbit.
Thus $w_{\theta,ji}=
\xi_j\frac{{M_i}^2}{M^2}\frac{h_{\rm orb}}{r_i}$,
where $h_{\rm orb}$ is the orbital specific angular momentum given by
${h_{\rm orb}}^2=r^4\omega^2=GMa(1-e^2)$ and
$\xi_1\equiv\chi_1(1-\mu_2)$, which yields
\begin{equation}
\label{vtw_ij}
(w_{\theta,12}-r_2\omega)=
r\omega\left(\frac{\xi_1{M_2}^2-{M_1}^2}{M_1 M}\right),
\end{equation}
and similar for $w_{\theta,21} - r_1\omega$.

\section{Rate of eccentricity change}
\label{sec2}
More useful than arbitrary variations of the physical parameters are their
rates of change in time, thus from this point on we will transform
$\delta\rightarrow {\rm d}/{\rm d}t$.
The distance and orbital velocity components are
\begin{equation}
\label{orbquants}
\begin{split}
\frac{a}{r}=&\frac{1+e\cos\theta}{1-e^2},\\
v_r=&\sqrt{\frac{G(M_1+M_2)}{a(1-e^2)}}e\sin\theta~~~{\rm and}\\
v_t=&\sqrt{\frac{G(M_1+M_2)}{a(1-e^2)}}(1+e\cos\theta),\\
\end{split}
\end{equation}
where $\theta$ is the orbital phase angle measured from the periastron.
It is clear from its dependence on the orbital phase-angle
that 
$\dot{e}$ 
is not constant along the orbit. This is true for constant mass loss
and accretion rates, but  
especially if mass loss and accretion rates also depend on
the orbital phase-angle. Employing the results we have
derived so far we calculate using Eq. (\ref{fve})
\begin{equation}
\label{e_theta}
\begin{split}
\dot{e}=&\frac{\dot{M}_{1,{\rm LOST}}(\theta)}{M}
\Bigg\{e+\cos\theta\Bigg\}\\
&+\frac{\dot{M}_{2,{\rm ACC}}(\theta)}{M_2}
\Bigg\{\frac{M_1}{M}
\Bigg[\left(\frac{\xi_1{M_2}^2}{{M_1}^2}-1\right)
(2\cos\theta+e+e\cos^2\theta)\\
&-e\sin^2\theta\Bigg]
+w_1\sqrt{\frac{a(1-e^2)}{GM}}\sin\theta\Bigg\}+\swapind.\\
\end{split}
\end{equation}
We note that the first term only depends on the assumption of
isotropic mass loss and it is the same as given by
\citet{2000A&A...357..557S} \citep[see also][]{2006epbm.book.....E}.
It shows that the eccentricity can
increase if the mass-loss rate is higher near periastron than
near apastron. The second term also depends on the
additional assumptions about the wind behaviour (see $\S$\ref{wbehave}).
These are complicated expressions that must be followed along the orbit
if they are to have a useful meaning. The orbital period is given by
$P=2\pi\sqrt{\frac{a^3}{G(M_1+M_2)}}$ and by defining the orbital average
of a parameter $A$ as $\left<A\right>\equiv\frac{1}{P}\int_0^{P}A{\rm d}t=
\frac{1}{P}\int_0^{2\pi}A\frac{{\rm d}t}{{\rm d}\theta}{\rm d}\theta$
we compute
the net rate of change of $a$ and $e$ in one orbital period.
When an orbital average is applied to 
Eq.
(\ref{e_theta}) we obtain 
an expression
which 
depends
on 
$\left<\dot{M}_{i,{\rm LOST}}\right>$, $\left<\dot{M}_{i,{\rm LOST}}\cos{\theta}\right>$,
$\left<\dot{M}_{i,{\rm ACC}}\right>$, $\left<\dot{M}_{i,{\rm ACC}}\cos{\theta}\right>$,
$\left<\dot{M}_{i,{\rm ACC}}\sin{\theta}\right>$, $\left<\dot{M}_{i,{\rm ACC}}\cos^2{\theta}\right>$
and $\left<\dot{M}_{i,{\rm ACC}}\sin^2{\theta}\right>$ for $i=1, 2$. These averages can be calculated
only after $\dot{M}_{i,{\rm LOST}}(\theta)$ and
$\dot{M}_{i,{\rm ACC}}(\theta)$ are known.

\subsection{Constant wind and constant mass accretion}
Usually the simplest case is assumed which corresponds to when
$\dot{M}_{i,{\rm ACC}}=\left<\dot{M}_{i,{\rm ACC}}(\theta)\right>$ and
$\dot{M}_{i,{\rm LOST}}=\left<\dot{M}_{i,{\rm LOST}}(\theta)\right>$ are
constant along the orbit \citep[e.g., ][]{2002MNRAS.329..897H}. This yields
\begin{equation}
\label{dedt_fix}
\frac{\left<\dot{e}\right>}{e}=
-\frac{\dot{M}_{2,{\rm ACC}}}{M_2}\left(
\frac{(1-\sqrt{1-e^2})}{e^2}\frac{(1-e^2)\xi_1{M_2}^2}
{M_1 M}\right) + \swapind.
\end{equation}
Clearly with the assumptions about the wind made here
the eccentricity can only decrease,
therefore if this effect is comparable to that of tidal
friction it will only help to circularise the orbit on a faster
timescale, otherwise it can be neglected
(\citealp{2002MNRAS.329..897H}\footnote{\citet{2002MNRAS.329..897H}
have a different expression for $\dot{e}/e$ due to different assumptions
made in their calculations, however, their expression and ours are of
the same order of magnitude.}).

\subsection{Accretion from a fast isotropic wind}
\label{acc_fw}
We model the mass accretion with the \citet{1944MNRAS.104..273B}
mechanism, in which
\begin{equation}
\label{BHacc}
\dot{M}_{2,{\rm ACC}}(\theta)=\alpha\frac{(GM_2)^2\dot{M}_{1,{\rm W}}}
{{r_A}^2{w_{{\rm rel},2}}^3w_1},
\end{equation}
where $\dot{M}_{1,{\rm W}}$ is the wind mass loss rate
from star $1$, $r_A$ the distance to the source
of the wind at the moment it was emitted and $\alpha$ is a
numerical factor which parameterises the efficiency of the accretion
mechanism.
In the case of a steady isotropic wind much faster than the
orbital motion 
the transverse component
of the wind velocity can be neglected, therefore one can safely assume that
$w_{{\rm rel},2}=w_1$ and $r_A=r$. Then the accretion rate
$\dot{M}_{2,{\rm ACC}}(\theta)=\dot{M}_{2,{\rm 0~ACC}}(1+e\cos{\theta})^2$,
where $\dot{M}_{2,{\rm 0~ACC}}=
\frac{\left<\dot{M}_{2,{\rm ACC}}(\theta)\right>}
{\left<(1+e\cos{\theta})^2\right>}$.
The mass from star $1$ which is lost from
the system is given by
$\dot{M}_{1,{\rm LOST}}=\dot{M}_{1,{\rm W}}-\dot{M}_{2,{\rm ACC}}$.
With this assumption the averaged rate of change of 
$e$ is
\begin{equation}
\label{dedt_fastw}
\begin{split}
\frac{\left<\dot{e}\right>}{e}=&\frac{\dot{M}_{2,{\rm 0~ACC}}}{M_2} \times\\
&\phantom{++}\frac{(1-e^2)^{3/2}}{2M_1(M_1+M_2)}\left(
3\xi_j{M_2}^2-2M_1(M_2+2M_1)\right) + \swapind.
\end{split}
\end{equation}
In contrast to Eq.(\ref{dedt_fix}), where $e$ always decreases
with time, Eq.(\ref{dedt_fastw}) shows that the eccentricity will
increase if $M_2/M_1>\left(1+\sqrt{1+12\xi_1}\right)/3\xi_1$
(e.g., $M_1\lesssim0.65M_2$
for the case of point masses, where $\xi_1=1$).

\subsection{Accretion from the slow wind of an AGB star}
\label{accAGB}
The case of an AGB star with an accreting companion is somewhat
different from the two cases we have already reviewed. The velocity
of the wind from an AGB star is of the order of the
stellar escape velocity, which is also of the order of the orbital
velocity, if the size of the AGB star is a significant fraction of
its Roche lobe. This implies that
$w_1\sim\sqrt{\frac{GM}{a(1-e^2)}}$,
so the transverse velocity of the wind cannot be neglected as in
the fast-wind case. Thus, making use of Eqs. (\ref{vtw_ij}) and
(\ref{orbquants}) we write
\begin{equation}
\begin{split}
&{w_{{\rm rel},2}}^2(\theta)=\left(\frac{GM}{a(1-e^2)}\right)\times\\
&\left[\left(\frac{w_1}{\sqrt{\frac{GM}{a(1-e^2)}}}
-\frac{M_1e\sin\theta}{M}\right)^2+
\left(\frac{[\xi_1{M_2}^2-{M_1}^2](1+e\cos{\theta})}{M_1M}\right)^2\right]
\end{split}
\end{equation}
and by means of Eq. (\ref{BHacc}) we express the accretion rate as
\begin{equation}
\dot{M}_{2,{\rm ACC}}(\theta)=
\frac{\dot{M}_{2,{\rm 0~ACC}}(1+e\cos{\theta})^2}
{\left[\left(\frac{w_1}{\sqrt{\frac{GM}{a(1-e^2)}}}
-\frac{M_1e\sin\theta}{M}\right)^2+
\left(\frac{[\xi_1{M_2}^2-{M_1}^2](1+e\cos{\theta})}{M_1M}\right)^2\right]^{3/2}},
\end{equation}
where $\dot{M}_{2,{\rm 0~ACC}}=
\left(\frac{GM}{a(1-e^2)}\right)^{3/2}
\left<\dot{M}_{2,{\rm ACC}}(\theta)\right>
\left<\frac{{w_{{\rm rel},2}}^3(\theta)}
{(1+e\cos{\theta})^2}\right>$ in this case. The orbital average can
be only calculated numerically, but we find that
when $w_1\gtrsim3\sqrt{\frac{GM}{a(1-e^2)}}$ the result of the
integration is well
approximated by the fast-wind assumption. The fact that the
orbital phase can change significantly between the moment the wind
was emitted and when it is accreted also introduces a
phase angle shift in the relation between $\dot{M}_{2,{\rm ACC}}$ and
$\dot{M}_{1,{\rm W}}$. However, we performed numerical
calculations of $\left<\dot{e}\right>/e$ by applying different phase
angle shifts in Eq. (\ref{BHacc}) and the
results do not differ by more than 15\% from the zero phase-angle case.

\subsection{Mass loss from an AGB star}
A non-AGB star has a steep density gradient in its
photospheric layers, i.e., it has a well defined radius. When
the star is smaller than its Roche lobe any mass loss is in the
form of a
wind which may be enhanced by the presence of the companion
\citep[e.g.,][]{1988MNRAS.231..823T}.
If such a star
becomes larger than its Roche lobe then the mass loss rate is
abruptly enhanced by orders of magnitude, governed by the
Roche lobe overflow mechanism. Thus there is an almost
discontinuous
transition from one regime to the other and they can be
treated as separate cases. AGB stars, on the other hand,
have a shallow surface density gradient
\citep[e.g., ][]{1988ApJ...329..299B,1989A&A...214..186P}
and  therefore
the transition between the wind and the Roche lobe overflow
mass-loss regimes is smooth. \citet{2005ApJ...623L.137K}
made observations in X-rays of the symbiotic binary system
Mira which show the AGB
star is surrounded by material in the shape of a lobe as predicted
by the Roche model. This suggests that the star is undergoing a
transitional form of mass loss that can be described as wind
Roche-lobe overflow.
This provides support to the idea
that the extended and weakly-bound atmosphere
of an AGB star can be
highly influenced by the tidal force exerted by the companion.
\citet{2001A&A...367..513F} computed the enhancement of the wind
mass-loss rate of giant stars in binary systems through the effect
of the Roche potential on the local surface temperature and gravity.
They find that the mass-loss enhancement depends on
$(R\subrats/R_{\rm L})^3$, where $R\subrats$ and $R_{\rm L}$
are radius of the AGB star and its Roche-lobe radius, respectively.
However, the overall effect is modest, less than a factor of 2,
unless $R\subrats$ is very close to $R_{\rm L}$. Their model does
not account for the wind Roche-lobe overflow transition expected
for AGB stars and suggested by the observations of Mira.
\citet{1988MNRAS.231..823T} propose that
the mass-loss rate of cool giants is enhanced by
the presence of a companion, driven by the tidal
friction torque. Thus, in their heuristic model the
mass-loss enhancement depends on the sixth power of the
$R\subrats/R_{\rm L}$ ratio.
Building on this idea, we propose the following AGB star
mass-loss model,
which provides a smooth transition between the single-star
wind mass-loss rate, $\dot{M}\subrats_{,{\rm w}}$, and the
Roche-lobe overflow mass-loss rate, $\dot{M}_{\rm RLOF}$:
\begin{equation}
\label{newML}
\dot{M}\subrats=
\dot{M}\subrats_{,{\rm w}}\left[
1+\left(\frac{\dot{M}_{\rm RLOF}}{\dot{M}\subrats_{,{\rm w}}}-1\right)
\left(\frac{R\subrats}{R_{\rm L}}\right)^6
\right].
\end{equation}

For eccentric orbits the Roche geometry does not apply so we utilise an
\emph{instantaneous} Roche-lobe which depends on the actual distance
between the stars, $r$, 
as
\begin{equation}
\label{rl}
R_{\rm L}(\theta)=\frac{a(1-e^2)}{(1+e\cos{\theta})}f(q\subrats),
\end{equation}
where $q\subrats$ is the mass ratio of the AGB star to its companion and
$f(q)=\frac{0.49q^{2/3}}{0.6q^{2/3}+\ln{\left(1+q^{1/3}\right)}}$,
following the approximation of \citet{1983ApJ...268..368E}.
With this prescription the mass loss rate along the orbit is not
constant, but enhanced when the system is at orbital phases close
to the periastron. Thus, if the AGB star does not fill its Roche
lobe in the periastron and there is no mass transfer
($\dot{M}_{i,{\rm 0~ACC}}=0$), the net contribution in one orbital period
to the eccentricity variation rate is
\begin{equation}
\label{edotmassloss}
\frac{\left<\dot{e}\right>}{e}=
\frac{\dot{M}_{\rm RLOF}-\dot{M}\subrats_{\rm,W}}{M}
\left(\frac{R\subrats}{af(q\subrats)}\right)^6
\frac{3(8+12e^2+e^4)}{8(1-e^2)^{9/2}}.
\end{equation}
Eq. (\ref{edotmassloss}) will always contribute to enhance
the eccentricity, given that
$\dot{M}_{\rm RLOF}$ is larger than $\dot{M}_{\rm AGB,W}$.
The effect of mass transfer to the companion will in many cases
act against the eccentricity enhancement, but
its contribution to Eq. (\ref{edotmassloss}) cannot be expressed
analytically and it has to be calculated numerically
(see $\S$\ref{acc_fw} and $\S$\ref{accAGB}). However, it becomes
significant only when a considerable amount of mass lost from the
AGB star is transferred.


\section{Comparison of timescales}
\label{sec3}
In this section we calculate and compare the timescales on which
the effects of circularisation and
eccentricity enhancement will take place,
$\tau_{\rm circ}$ and $\tau_{\rm e}$ respectively.
\citet{2002MNRAS.329..897H} give the following expression for the
tidal evolution of $e$ due to convective damping by combining the
results of
\citet[][see also \citealt{1989A&A...223..112Z}]{1977A&A....57..383Z}
and \citet{1981A&A....99..126H}:
\begin{equation}
\label{taucirc}
\begin{split}
\frac{1}{\tau_{\rm circ}}\equiv-\frac{\dot{e}}{e}=&
\frac{18}{7}\frac{f_{\rm conv}}{\tau_{\rm conv}}\frac{M_{\rm env}}{M\subrats}
\frac{(1+q\subrats)}{q\subrats^2(1-e^2)^{13/2}}
\left(\frac{R\subrats}{a}\right)^8\\
&\times\left[f_3(e^2)-\frac{11}{18}(1-e^2)^{3/2}f_4(e^2)
\frac{\Omega\subrats_{\rm ,s}}{n}\right],
\end{split}
\end{equation}
where $M_{\rm env}$ is the envelope mass of the AGB star and
$\tau_{\rm conv}$ is the convective turnover timescale.
A good approximation to the depth of the convective zone in an AGB star
is its radius, thus by employing Eq. (31) from
\citet{2002MNRAS.329..897H} we obtain $\tau_{\rm conv}=0.2351
(M_{\rm env}{R\subrats}^2/L\subrats)^{1/3}{\rm yr}$,
with $L\subrats$ the luminosity of the star and the parameters
expressed in solar units. 
$\Omega\subrats_{\rm ,s}/n$ is the ratio of spin angular rotation
rate of the AGB star, $\Omega\subrats_{\rm ,s}$,
to the average orbital angular velocity, $n$.
The functions $f_x(e^2)$ are those derived by
\citet{1981A&A....99..126H} and the numerical factor
$f_{\rm conv}=f'\min\left(1,(P_{\rm tid}/2\tau_{\rm conv})^2\right)$
is from \citet{1996ApJ...470.1187R}, where
${P_{\rm tid}}^{-1}=|n-\Omega\subrats_{\rm ,s}|/2\pi$.
\citet{1995A&A...296..709V} found from an analysis of observed
red-giant binaries that $f'\approx1$, however,
this value may be different for the case of AGB
stars.
It is important to note
that this calculation of $\tau_{\rm circ}$ gives a stronger
circularising effect for orbits with finite eccentricity values
than the usually employed prescription by
\citet{1977A&A....57..383Z} which assumes
an almost circular orbit
\citep[e.g.,][]{1996ApJ...470.1187R,2000A&A...357..557S}.

We estimate the Roche-lobe overflow mass-loss rate
$\dot{M}_{\rm RLOF}$ following the
prescription of \citet[][see also \citealt{1983A&A...121...29M,1989A&A...214..186P}]{1988A&A...202...93R}, where
\begin{equation}
\label{mrlof}
\dot{M}_{\rm RLOF}=\frac{2\pi}{\sqrt{e}}
\left(\frac{k_{\rm B}}{m_{\rm H}\mu\subrats}T\subrats\right)^{3/2}
\frac{{R\subrats}^3}{GM\subrats}\rho\subrats_{\rm ,ph}F(q\subrats^{-1}),
\end{equation}
with $F(x)=1.23+0.5\log{x}$. The photospheric
density of the AGB star is given by
\citep{1990sse..book.....K,2006NewA...11..396S}
\begin{equation}
\rho\subrats_{\rm ,ph}=
\frac{2}{3}\frac{m_{\rm H}\mu\subrats}{k_{\rm B}}
\frac{GM\subrats}{R\subrats^2\kappa\subrats T\subrats},
\end{equation}
where $\kappa\subrats$ is the opacity of the envelope of the AGB star
and $\mu\subrats$ and $T\subrats$ are the mean molecular weight and
effective temperature of the stellar surface, respectively.
$m_{\rm H}$ is the mass of the hydrogen atom and and $k_{\rm B}$ is the
Boltzmann constant.
A typical 1.5-2.5 M$_{\odot}$ AGB star has a surface temperature of about
3300K, a radius of about 300 R$_{\odot}$ and a luminosity of about 10000
L$_{\odot}$. Most of the surface material in such a star is weakly
ionised due to the low surface temperature, so we assume a typical
surface molecular weight of 1.3
\citep[see e.g.,][]{2006NewA...11..396S}.
According to the envelope opacity calculations of \citet{2002A&A...387..507M},
the corresponding opacity at this temperature is approximately
$\kappa\subrats=2.5\times10^{-4}{\rm ~cm^2~g^{-1}}$
for densities of the order of 
$10^{-13}$ -- $10^{-11}$ $\mathrm{g~cm^{-3}}$,
which is the expected range of AGB surface densities. 
Employing this 
value of $\kappa\subrats$ and the luminosity, radius, mass and
mean molecular weight mentioned
above, the prescription of Eq. (\ref{mrlof}) yields
$\dot{M}_{\rm RLOF}\sim10^{-3}$ M$_{\odot}~yr^{-1}$. 
This is about 10000-30000 times larger than the single-star
wind mass-loss rate of our typical star at the beginning of the AGB
lifetime, calculated with the prescription
of \citet{1993ApJ...413..641V}.
By the end of the AGB phase, at the onset of the superwind phase,
$\dot{M}_{\rm RLOF}$ is still 50-200 times larger than the wind mass-loss rate.

\subsection{Eccentricity of the Sirius system}
\label{sec:sirius}
Sirius A, a main sequence star of 2.1 M$_{\odot}$, and Sirius B, a
white dwarf of 1.05 M$_{\odot}$ \citep{1978ApJ...225..191G},
form a binary system with period $P=50.1$ yr and eccentricity
$e=0.59$ \citep{1960JO.....43..145V}. Given the semi-major axis of
the orbit, $a=20$ AU, and the eccentricity of this system,
the distance between
its components at periastron is about 1700 R$_{\odot}$.
By the time the primary star (initially a star of about 5.5-6 M$_{\odot}$)
was in the AGB phase its radius was so large ($R\approx750$ R$_{\odot}$)
that it could have
filled its Roche lobe during periastron passages. Whether or not
the Roche lobe was filled, a strong tidal
interaction must have occurred due to the proximity of the
system components. We have tracked the evolution of binary systems which
evolve into a  main
sequence star with the mass of Sirius A and a white dwarf with the
mass of Sirius B, which from here on we will call
Sirius-like systems.
To compute the binary evolution models we employed a version
of the rapid synthetic code of binary evolution of
\citet{2006A&A...460..565I}, which makes use of modifications
described  in \citet{2007A&A...469.1013B}.
The code employs Eq. \ref{taucirc} to model the effect of
circularisation.
The set of initial conditions for the models consists of the
following:
\begin{itemize}
\item A primary with initial mass of 5.7 M$_{\odot}$, which ends its life
as a white dwarf similar to Sirius B, independently of the choice
of mass loss.
\item 2 initial secondary masses (2.00 and 2.05 M$_{\odot}$)
which yield a star similar to Sirius A.
\item 25 initial separations, $a_{\rm i}$, logarithmically
separated between 10 AU and 500 AU. 
\item 25 initial eccentricities, separated linearly
between 0.5 and 0.99.
\item Solar metallicity, $Z=0.02$.
\end{itemize}

   \begin{figure}
   \centering
   \includegraphics[width=0.5\textwidth]{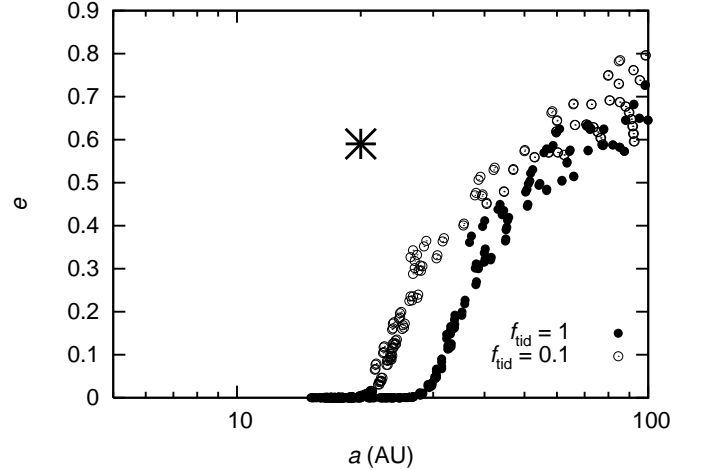}
   \caption{Computed eccentricity of Sirius-like binary models
     without any eccentricity-enhancing mechanism
     as function of their separation,  assuming the
     standard binary interaction. The filled circles
     indicate the orbital parameters of Sirius-like systems computed
     by assuming a tidal dissipation factor $f'=1$.
     The open circles are similar, but assuming that the tidal
     dissipation is ten times weaker. The star indicates the observed
     orbital parameters of the Sirius system, which has a notably
     higher eccentricity than what standard models predict for its
     separation, even with substantially reduced tidal friction.
   }
      \label{fig:sirius_std}
   \end{figure}

In the standard scenario of tidal dissipation with constant mass loss
along the orbit the eccentricity will always decrease.
Fig. \ref{fig:sirius_std} shows the distribution of the synthetic
Sirius-like systems in the $a$-$e$ plane compared to the observed
position of the Sirius system in this plane.
The solid circles indicate the orbital properties of Sirius-like
systems when assuming $f'=1$, which show that the 
standard scenario of tidal interaction in binary evolution
predicts complete circularisation for systems separated up to
about 30 AU. The closest system with an eccentricity of
0.59 has a separation of about 55 AU.
The open circles show the orbital parameters predicted for
Sirius-like systems when the strength of tidal dissipation is
reduced to $f'=0.1$. The results are still far
from reproducing the observed Sirius system. A separation of
at least about 40 AU is required to obtain the observed
eccentricity of 0.59.
To reproduce the eccentricity and separation of Sirius within this
scenario a tidal dissipation factor
$f'<10^{-3}$ is needed, which is too strong a reduction to be acceptable.
This demonstrates that the standard scenario of binary
evolution cannot explain the observed eccentricity of the
Sirius system, because it predicts that the system should be
circularised.


However, this is not the case if the tidally enhanced
mass-loss we propose in Eq. (\ref{newML}) is considered. Due to
its variable mass loss along the orbit an eccentricity enhancing effect
counteracts the tidal circularisation.
We compare $\tau_{\rm e}$ and $\tau_{\rm circ}$ to see which effect
is dominant.
We estimate their values at the moment of evolution when
the tidal dissipation effects are strongest, which is the bottleneck
for ending up with an eccentric orbit. This occurs when the
AGB star reaches its maximum radius, $700$ R$_{\odot}$, at which point
it has a luminosity of $38000$ L$_{\odot}$ and a core and envelope mass
of $1.0$ M$_{\odot}$ and $3.5$ M$_{\odot}$ respectively. For the
calculation of the Roche-lobe overflow mass-loss rate we assume
$\kappa\subrats=10^{-4}$, which approximates the values
calculated by \citet{2002A&A...387..507M} for the envelope opacity
of high-metallicity stars with a temperature of about 3000 K.
This gives $\dot{M}_{\rm RLOF}=1.944\times10^{-3}$ M$_{\odot}~yr^{-1}$
according to Eq. \ref{mrlof}, while the wind mass-loss rate at this point is
$\dot{M}\subrats_{\rm,W}=5.168\times10^{-5}$ M$_{\odot}~yr^{-1}$.
While mass loss from a system tends to increase its separation,
the tidal interaction tends to (pseudo-)synchronise the rotation
of the AGB star with the orbit and thereby shrink the orbit
due to transfer of orbital angular momentum.
This orbital shrinking effect is especially strong in the beginning of the
AGB phase, when the star expands rapidly and rotates slowly.
However,
once the AGB star rotation approaches a
(pseudo-)synchronised rate with the orbit,
the widening of the orbit due to mass loss dominates over the
effect of tidal shrinking.
Assuming this last regime,
%
   \begin{figure}
   \centering
   \includegraphics[width=0.5\textwidth]{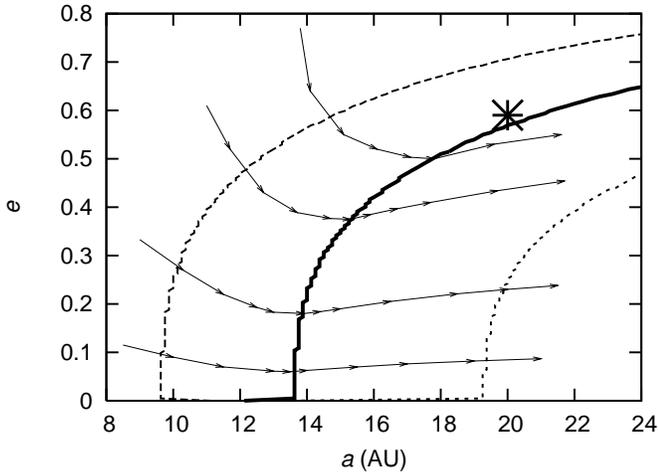}
      \caption{Behaviour of eccentricity of a binary system with
	a $4.5$M$_{\odot}$ AGB primary (initially a $5.5$M$_{\odot}$
	star) and a $2.1$M$_{\odot}$ MS secondary as the system widens
	due to mass loss.
	The thick solid line indicates $\tau_{\rm e}=\tau_{\rm circ}$
	when $f'$ is assumed to be 0.5 and 5\% of the mass lost via the
	enhanced mass-loss mechanism is accreted by the companion.
	The arrows indicate instantaneous directions in the diagram
	which a binary system will follow as it evolves.
	The star represents the current position in the
	$a$-$e$ plane of the Sirius system.
	The thin dashed and dotted lines also indicate
	$\tau_{\rm e}=\tau_{\rm circ}$, but considering
	$f'=0.25$ and $f'=1$, respectively.
	}
      \label{fig_sirius_scheme}
   \end{figure}
Fig. \ref{fig_sirius_scheme} shows how the $a$-$e$ plane
is divided into a region where the eccentricity decreases (upper-left area)
and a region where the eccentricity is enhanced (lower-right area).
Thus, if we adopt a modest reduction of the tidal dissipation strength
($f'=0.5$),
any system with a separation larger than about 14 AU will eventually
cross to the eccentricity-enhancing region, avoiding a circular orbit
by the end of the AGB phase. In particular, 
the orbit of a system with orbital parameters similar to Sirius will
remain eccentric.
The thin dotted and dashed lines in
Fig. \ref{fig_sirius_scheme} show that the
location of the dividing line between decreasing and increasing
eccentricity is sensitive to the tidal strength factor $f'$. This
is because the ratio of timescales $\tau_{\rm e}/\tau_{\rm circ}$
only varies as $(R\subrats/a)^2$, i.e., a much weaker dependence
than either of the individual timescales. There is a similar
sensitivity to the adopted average opacity $\kappa\subrats$ through
$\dot{M}_{\rm RLOF}$, which is also not certain, given that
Eq. \ref{mrlof} is only approximate. Given these uncertainties, the results
are not definitive. However, we have shown that
by adopting reasonable values for the uncertain parameters the orbit
of Sirius can be kept significantly eccentric. Finally we note that
the results are insensitive to the amount of mass accreted by the
companion, as long as this is less than about 10-15\%
of the mass lost by the AGB star.

   \begin{figure}
   \centering
   \includegraphics[width=0.5\textwidth]{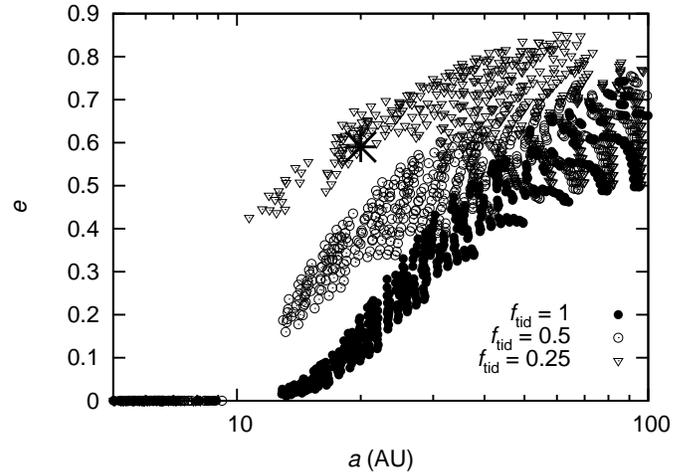}
   \caption{Computed eccentricity of Sirius-like binary
     models which take into account our proposed eccentricity
     enhancing mechanism plotted as function of their
     separation. The filled circles indicate the orbital
     parameters of Sirius-like systems computed by assuming
     a tidal dissipation factor $f'=1$. The open circles
     and triangles indicate the results for similar calculations,
     but assuming that the tidal dissipation factor $f'$ is 0.5 and
     0.25, respectively. The star indicates the observed orbital
     parameters of the Sirius system, which are well reproduced if
     $f'=0.25$ is assumed.}
      \label{fig:sirius_pumped}
   \end{figure}

So far we have shown how our eccentricity-enhancing mechanism works by
employing values of single-star models for the
AGB primary at a given moment of evolution. However,
the overall effect on the orbit of a given binary depends on how
the relevant parameters change as a function of time as the binary
evolves from zero age to its current state.
We have therefore implemented the orbital evolution formulae
derived in Sect. \ref{sec2} into the synthetic binary evolution code of
\citet{2006A&A...460..565I}, including our enhanced mass-loss prescription.
Full details of this implementation are given in
\citep{BMP07}. We calculate models for the same set of initial conditions as
described above.
The choice of the amount of mass that the companion accretes does not
affect our results significantly, unless it is larger than about 15\%
of the total mass lost by the primary.
We assume that 5\% of the enhanced mass loss
($\dot{M}\subrats-\dot{M}\subrats_{,{\rm w}}$) is accreted by the companion,
which increases the final mass of the secondary to that of Sirius A.
Fig. \ref{fig:sirius_pumped} shows a comparison of the
synthetic Sirius-like systems resulting from these models to the
observed orbital properties of Sirius. The filled circles
show the eccentricity and separation predicted by models which
assume the standard tidal dissipation, $f'=1$.
Compared to the case when only tidal dissipation is considered
(Fig. \ref{fig:sirius_std}), the resulting binaries are
significantly more eccentric in the separation range 15-40 AU,
and the effect is comparable to reducing the tidal strength to
$f'=0.1$ in Fig. \ref{fig:sirius_std}.
The triangles in
Fig. \ref{fig:sirius_pumped} show that to reproduce the orbital
properties of Sirius the tidal dissipation has to be weakened
by only a factor of 4, which is still within a range of reasonable
values, as opposed to a reduction by more than a factor 1000 needed
when only tidal circularisation is include.

\subsection{Eccentricities of the barium stars}
\begin{figure}
   \centering
   \includegraphics[width=0.5\textwidth]{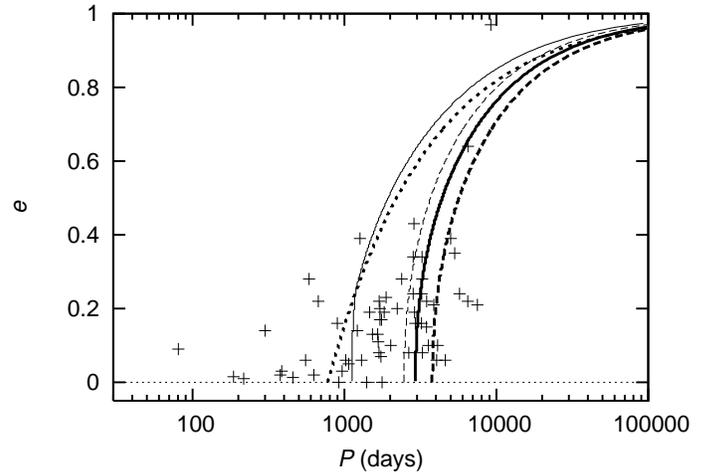}
      \caption{Distribution in the $P$-$e$ plane of barium star
	and the limits for circularisation
	set by different models.
	The crosses indicate the observational data of
	barium stars form \citet{1998A&A...332..877J}.
	The thick lines delimit the regions where eccentric
	and circularised orbits 
	are expected and are calculated for a system as
	described in the text, by assuming a tidal dissipation factor
	$f'=1$. The thin lines show the same limits as the thick lines,
	but this time calculated by assuming $f'=0.1$.
	Solid lines indicate the circularisation limit when the
	eccentricity pumping mechanism we propose is considered.
	The dashed lines show the circularisation limit in the
	standard scenario. The dotted line indicates the limit for
	which Roche lobe overflow occurs at the periastron and the
	horizontal thin dotted line indicates $e=0$.
      }
      \label{e_P_ba}
   \end{figure}

Barium stars show over-abundances of $s$-process elements, most
prominently of barium, and are found in binary systems with white
dwarf companions. This constitutes evidence for
mass accretion from their companions when the latter were asymptotic
giant branch (AGB) stars, given that $s$-process elements are
synthesised in the AGB phase of evolution. Barium star binary systems
are observed to have periods between 80 and 10000 days and most of
the systems with periods larger than about 600 days are significantly
eccentric (see Fig. \ref{e_P_ba}).
%
%
%
These systems pose a problem to the
standard binary evolution scenario which predicts that
orbits with periods shorter than about 3500-4000 days should
have been circularised due to the tidal dissipation that must have taken
place when the primary was an AGB star \citep[e.g.,][]{2003ASPC..303..290P}.
The eccentricity enhancement
resulting from our proposed AGB mass loss provides a mechanism
which can effectively compete with tidal circularisation.
We calculate the relevant timescales for a binary system with a
1.5 M$_{\odot}$ AGB primary and a 1.1 M$_{\odot}$ secondary, with
metallicity $Z=0.008$. We assume that 10\% of
the mass lost by the AGB star is accreted by the companion,
so that the latter accretes enough material to enhance
its $s$-process element abundances significantly and become
a barium star.
The calculations are made at
the AGB stage of evolution where the tidal dissipation is 
strongest (the circularisation bottleneck), where the radius of the
AGB star is $280$ R$_{\odot}$, its luminosity is $7500$ L$_{\odot}$ and
its core and envelope masses are $0.64$ M$_{\odot}$ and $0.6$ M$_{\odot}$
respectively.
Applying these values to Eq. \ref{mrlof} we obtain that
$\dot{M}_{\rm RLOF}=1.75\times10^{-3}$ M$_{\odot}~yr^{-1}$
according to Eq. \ref{mrlof}, while the wind mass-loss rate at
this point is
$\dot{M}\subrats_{\rm,W}=1.02\times10^{-5}$ M$_{\odot}~yr^{-1}$.

Our results are shown in Fig. \ref{e_P_ba} and are compared to the
observed data from \citet{1998A&A...332..877J} in the $P$-$e$ plane.
The thick dashed line separates the region where the orbits will be
circularised (to the left of the line) from the region where eccentric
orbits are possible (to the right of the line) in the standard binary
scenario, where only the tidal circularisation is acting. This limit
is calculated by assuming $\tau_{\rm circ}^{-1}=M/\dot{M}$
and a tidal dissipation factor $f'=1$, and reasonably well reproduces
the limit resulting from binary population synthesis
calculations \citep[e.g.,][]{2003ASPC..303..290P}.
The majority of the systems cannot be
explained in this picture. The thick solid line indicates the
same limit, but with the eccentricity pumping
mechanism included, i.e., when $|\tau_{\rm e}^{-1}-\tau_{\rm circ}^{-1}|=M/\dot{M}$.
The evolution of binary systems in the $P$-$e$ plane is similar to that
in the $a$-$e$ plane (shown in Fig. \ref{fig_sirius_scheme}), thus the
eccentricities of the systems to the right of the thick line
can be explained. This means that
with the default tidal strength the eccentricity-enhancement mechanism
cannot compete effectively with the tidal circularisation, except for
a very limited period range around 3000 days.
However, the tidal dissipation factor $f'$ is not very certain for
AGB stars, given that it has been measured to be approximately unity
mostly using  observations of red giants \citep{1995A&A...296..709V}.
Hence it is reasonable to calculate the
circularisation limits for smaller values of $f'$, as our
calculations for the Sirius system suggest. Fig.
\ref{e_P_ba} shows in thin lines the same limits that we have already
described above, but assuming $f'=0.1$. 
In this case the eccentricity of almost all
systems which do not fill their Roche-lobes can be explained.
\citet{2000MNRAS.316..689K} found that with a similar choice of $f'$
they could reproduce the eccentricities of all observed barium stars,
including those with periods less than 1000~days. In their simulations
they included a tidally enhanced AGB mass loss rate as well, but
without any eccentricity-enhancing mechanism to compete against tidal
circularisation. However, the results they report are difficult to
understand in the light of our findings, especially because they
excluded all systems in which the AGB star filled its Roche lobe.
The systems to the left of the thick dotted line in Fig. \ref{e_P_ba}
fill their Roche lobes during the AGB
and in most cases enter a common envelope (CE)
phase. The evolution of the orbit during this phase is not a well understood
problem and is also beyond the scope of this study. However,
it seems more likely that an already eccentric system
which enters a CE phase may remain eccentric than that a circular system
becomes eccentric during the CE phase. Thus a mechanism which at least
allows orbits to remain eccentric until the point when Roche-lobe overflow
occurs may be necessary to explain the eccentric systems which are to the
left of the Roche-lobe limit. As shown in Fig. \ref{e_P_ba}, assuming
weak tidal friction with $f'=0.1$ makes it possible that systems with an
AGB primary fill
their Roche lobe with a significant eccentricity.

It is important to note that the reduction of the
tidal strength needed to explain the eccentricities
of most observed barium-star systems which did not fill their
Roche lobe ($f'=0.1$)
is calculated for one specific system, assumed to be
representative. 
In order to test whether the eccentricity-enhancing mechanism
explains the properties of the barium stars,
the full evolutionary history has to be followed
(as in the case of Sirius) for a variety of system parameters
in a binary population synthesis study.
Furthermore, not only the orbital parameters
have to be reproduced, but also their abundance distributions.
Such an analysis is beyond the scope of this work and we
will present it in a subsequent paper.
Preliminary results of population synthesis of barium stars
indicate that $f'=0.25$
is needed for reproducing the eccentricities
of most observed systems which have not filled their Roche lobes,
which is consistent with the reduction of tidal dissipation strength
needed to reproduce the orbit of the Sirius system.

\section{Conclusions}
\label{sec4}
We have revised the description of the evolution of orbital
parameters due to mass loss
and mass transfer in a binary system where the primary is an AGB star.
We also propose a tidally enhanced mass-loss rate for AGB stars in
binary systems which allows a smooth transition between wind mass loss
and mass loss due to Roche lobe overflow. With our revised prescription
for the evolution of the orbital eccentricity and the fact that our
proposed mass loss is not constant along an eccentric orbit,
we find an eccentricity enhancing
mechanism which 
counteracts the
circularising effect of tidal dissipation. 
We have shown that the standard scenario of binary interaction
cannot explain the observed orbital parameters of the eccentric binary
system Sirius unless the strength of tidal dissipation in AGB stars
is at least 3 orders of magnitude smaller than in normal red giants.
On the other hand,
our models with the eccentricity-enhancing mechanism can
reproduce the orbital properties of the Sirius system with a
reasonable choice of the tidal dissipation parameter ($f'=0.25$).
Our eccentricity enhancing mechanism also allows binary
systems containing barium stars
to remain eccentric with periods as short as about 1000 days under
reasonable parameter assumptions, while under the same assumptions
in the standard scenario
of tidal circularisation only systems with periods longer than
about 2500 days are expected to show a significant eccentricity.
We also show that we can explain the eccentricities of almost all
barium star systems which
do not fill their Roche lobes if the uncertain convective tidal
dissipation strength in AGB stars
is reduced by a factor of 10, compared to what is measured for red
giant stars. Moreover, this assumption allows
Roche-lobe overflow to occur in significantly eccentric systems,
which may be the key to explain shorter-period barium-star
systems which are still eccentric.


Whether our eccentricity-enhancing mechanism is a viable solution
to the problem of explaining short-period eccentric
binaries with (former) AGB stars,
in particular barium stars, needs to be tested with a binary population
synthesis model. Results of such a test will be given in a separate paper
\citep{BMP07}. 
Preliminary barium star population synthesis calculations indicate that
when the full evolutionary history of a binary population is taken
into account, a reduction of the tidal dissipation strength of only 
$f'=0.25$ is needed to reproduce the observed systems.
This is consistent with our findings about the values of $f'$ needed
to reproduce the Sirius system.


\begin{acknowledgements}
The authors would like to thank Frank Verbunt for pointing
out that the problem of the eccentricity of the Sirius system is
similar to that of the eccentricities of the barium stars.
\end{acknowledgements}

\bibliography{refs}        

\begin{appendix}
\section{Derivation of variation of orbital parameters}
\label{appa}
\begin{figure}
   \includegraphics[width=0.25\textwidth]{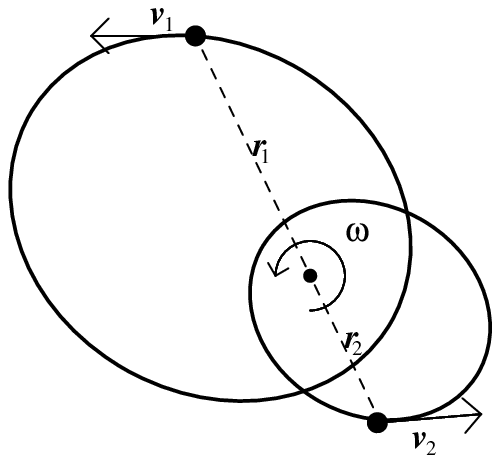}%
   \includegraphics[width=0.25\textwidth]{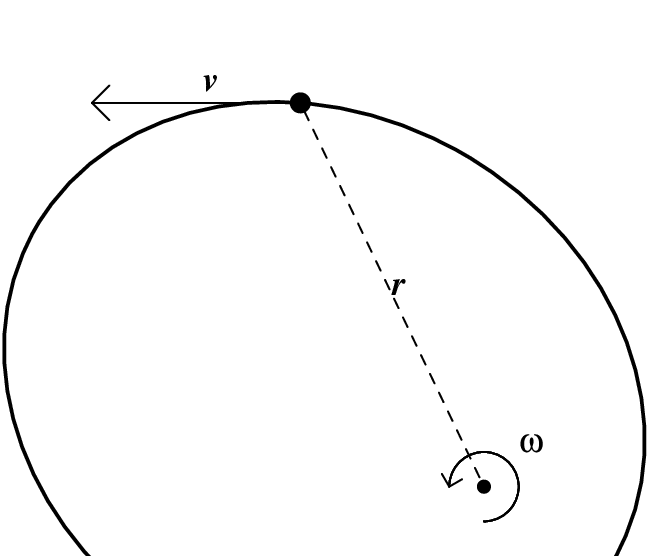}
   \caption{The geometric relation between the positions and velocities of
   both components in a binary system (left) and the equivalent reduced
   mass problem (right).}
   \label{fig:orbit}
\end{figure}
The orbit of a binary star system can be described by the eccentricity $e$ and
the  semi-major axis $a$. The change in these orbital parameters can be related
to changes in the masses $M_{1,2}$ of the two stars, either due to mass loss or
mass accretion, through the conservation laws for the total orbital energy 
$E_\mathrm{orb}$ and the orbital angular momentum $J_\mathrm{orb}$, given by
\beq
E_\mathrm{orb} = - \frac{GMm}{2a},
\eeq
and
\beq\label{eqn:j_orbit}
J_\mathrm{orb}^2 = m^2 r^4 \omega^2 = GMm^2 a (1-e^2).
\eeq
Here $\omega$ is the instantaneous angular velocity, $r$ is the
instantaneous separation among the system components, $M = M_1 + M_2$ 
is the total mass and $m = M_1 M_2 / M$ is the reduced mass of the system.
The variations of the orbital parameters $a$ and $e$ are then given by
\beq\label{eqn:var_a}
\frac{\delta a}{a} =
\frac{\delta M}{M} +
\frac{\delta m}{m} -
\frac{\delta E_\mathrm{orb}}{E_\mathrm{orb}}
\eeq
and
\beq
\frac{\delta(1-e^2)}{1-e^2} =
2 \frac{\delta J_\mathrm{orb}}{J_\mathrm{orb}} +
  \frac{\delta E_\mathrm{orb}}{E_\mathrm{orb}} -
2 \frac{\delta M}{M} -
3 \frac{\delta m}{m}.
\eeq
In the following we will derive expressions for $\delta E_\mathrm{orb}$ and
$\delta J_\mathrm{orb}$.

\subsection{Variation of the Orbital Energy}
The total orbital energy is the sum of the kinetic energy and the gravitational
energy,
\beq\label{eqn:e_orbit}
   E_\mathrm{orb} = 
   \frac{1}{2}m v^2 
   - 
   \frac{G M_1 M_2}{r},
\eeq
where $r = \left| \vec{r}_1 - \vec{r}_2 \right|$ is the instantaneous distance
between the two stars and $\vec{v} = \vec{v}_1 - \vec{v}_2$ is their
relative velocity (see Fig. \ref{fig:orbit}). The variation of (\ref{eqn:e_orbit})
is given by
\beq\label{eqn:var_eorb}
   \delta E_\mathrm{orb} = 
      \frac{1}{2} m v^2 \left(
         \frac{\delta m}{m} +
         \frac{\delta (v^2)}{v^2}
      \right)
      -
      \frac{G M m}{r} \left(
         \frac{\delta M}{M} +
         \frac{\delta m}{m} -
         \frac{\delta r}{r}     
      \right).
\eeq

The velocity $\vec v$ has components $v_r$ and $v_\theta = r\omega$ in
polar coordinates, so that the variation of $\delta (v^2)$ can be written
as
\beq
\delta(v^2) = 2 v_r \delta v_r + 
   2 r\omega \left(r \delta\omega + \omega \delta r\right).
\eeq
For later use, we will need to know the variation in terms of the
variations for the individual components,
\beq\label{eqn:vsquared}
   \delta(v^2) = 2 v_r \delta v_{r,1} + 
      2 r\omega (r_1 \delta\omega + \omega \delta r_1) + \swapind.
\eeq
The notation $ + \swapind$ means:
add terms to the left again with 
indexes $1$ and $2$ interchanged. Note that this is an addition despite
$\vec{v}$ being defined as the difference $\vec{v}_1 - \vec{v}_2$. The reason
is that contrary to a Cartesian basis the direction of basis vectors in 
polar coordinates is not fixed but depends on the angular coordinate $\theta$.

We will consider two effects that change the mass of each star: Bondi-Hoyle
accretion \citep{1944MNRAS.104..273B} and isotropic mass loss. The latter
does not affect the velocity of the star relative to the centre of mass,
but the former does. Let $\delta M_1 = \delta M_{1,\mathrm{ACC}}$ denote the 
mass accreted by star 1 and let $\vec{w}_{21}$ be the velocity of the wind of 
star 2 at the moment it is accreted by star 1 and $\vec{w}_{12}$ vice versa. 
Then conservation of momentum gives
\beq\label{eqn:momentum_conservation}
   M_1 \delta \vec{v}_1 = 
      \left( \vec{w}_{21} - \vec{v}_1 \right)\delta M_{1,\mathrm{ACC}}
\eeq
and similar for star 2. Inserting this in (\ref{eqn:vsquared}) gives the
relation
\beq
   \delta(v^2) = 2 \vec{v}
      \cdot
      \left( \vec{w}_{21} - \vec{v}_{1} \right)\frac{\delta M_{1,\mathrm{ACC}}}{M_1}
      + \swapind.
\eeq
It now remains to express $\delta M$ and $\delta m$ in terms of the variation
of the individual masses $\delta M_1$ and $\delta M_2$. For the purpose of
bookkeeping we split the mass lost from star 1 (and similarly for star 2) into 
two parts: the amount $\delta M_\mathrm{1, LOST}$ that is lost from the system
and the amount $\delta M_\mathrm{2, ACC}$ that is eventually accreted by star 2.
Then
\beq\label{eqn:var_dm1}
\delta M_1 = 
- \delta M_{1,\mathrm{LOST}}
- \delta M_{2,\mathrm{ACC}} 
+ \delta M_{1,\mathrm{ACC}}
\eeq
and
\beq\label{eqn:var_dm2}
\delta M_2 = 
- \delta M_{2,\mathrm{LOST}}
- \delta M_{1,\mathrm{ACC}} 
+ \delta M_{2,\mathrm{ACC}}.
\eeq
The variation of the total mass $\delta M$ is just $\delta M_1 + \delta M_2$
while the variation of the reduced mass is
\beq\label{eqn:var_dmr}
\frac{\delta m}{m} =
-\frac{\delta M_{1,\mathrm{LOST}}}{M_1} \frac{M_2}{M_1+M_2}
+\frac{\delta M_{1,\mathrm{ACC}}}{M_1} \left(1 - \frac{M_1}{M_2}\right)
+ \swapind.
\eeq
If the transfer of momentum due to the wind is instantaneous
the distance $r$ between the two stars is not changed, hence
$\delta r = 0$. Inserting (\ref{eqn:vsquared}), (\ref{eqn:var_dm1}),
(\ref{eqn:var_dm2}) and (\ref{eqn:var_dmr}) into (\ref{eqn:var_eorb}) then
gives
\beq\label{eqn:var_eorb_final}
\begin{split}
\frac{\delta E_\mathrm{orb}}{E_\mathrm{orb}} =
& \frac{\delta M_{1,\mathrm{ACC}}}{M_1}\left[
   1-\frac{M_1}{M_2} - \frac{2a}{G(M_1+M_2)} \vec{v}\cdot (\vec{w}_{21}-\vec{v}_1)
\right]
\\
& - \frac{\delta M_{1,\mathrm{LOST}}}{M_1} \left[
   1 + \left(\frac{2a}{r}-1\right) \frac{M_1}{M_1+M_2}
\right]
+(1 \Leftrightarrow 2)
\end{split}
\eeq
which is the same as (\ref{fvEvect}).

\subsection{Variation of the Orbital Angular Momentum}
The variation of the orbital angular momentum (\ref{eqn:j_orbit}) is
\beq
\frac{\delta J_\mathrm{orb}}{J_\mathrm{orb}} = 
\frac{\delta m}{m} + \frac{\delta r}{r} + \frac{\delta v_\theta}{v_\theta},
\eeq
where as before $v_\theta = r \omega$. Using (\ref{eqn:momentum_conservation}),
(\ref{eqn:var_dmr}) and the fact that $r$ does not depend on the variation
of the two masses immediately gives
\beq
\begin{split}
\frac{\delta J_\mathrm{orb}}{J_\mathrm{orb}} =
& \frac{\delta M_{1,\mathrm{ACC}}}{M_1} \left(
   1 - \frac{M_1}{M_2}
   +\frac{{w}_{\theta,21} - r_1\omega}{r_1\omega}
\right)\\
& -\frac{\delta M_{1,\mathrm{LOST}}}{M_1} \frac{M_2}{M_1 + M_2}
+ \swapind
\end{split}
\eeq
which is the same as (\ref{fvJorbvect}).

\subsection{The Variation of the Semi-Major Axis}
The variation of the eccentricity that results from the above considerations
is given in (\ref{e_theta}) in the text. For completeness we here give the
corresponding expression for the variation in $a$. Inserting
(\ref{eqn:var_eorb_final}) and (\ref{eqn:var_dmr}) in (\ref{eqn:var_a}) we
obtain
\beq
\begin{split}
\frac{\delta a}{a} 
= &
\frac{\delta M_{1,\mathrm{ACC}}}{M_1} 
\frac{2a}{G(M_1+M_2)} \vec{v}\cdot (\vec{w}_{21}-\vec{v}_1)
+\\
&+
\frac{\delta M_{1,\mathrm{LOST}}}{M_1}
\frac{M_1}{M_1+M_2}\left(\frac{2a}{r}-1\right)
+
(1 \Leftrightarrow 2).
\end{split}
\eeq
\end{appendix}
\end{document}